\newcommand{\shorttitle}{Hydrogen Plasmas in Strong Magnetic Fields}
\newcommand{\shortauthor}{A.Y. Potekhin et al.} 
\newcommand{\contnum}{1} 
\newcommand{\contvolum}{39} 
\newcommand{\contyear}{1999} 
\newcommand{\contpages}{in press} 
\documentstyle[twoside,epsfig,amssymb]{cpp}
\newcommand{\beq}{\begin{equation}}
\newcommand{\eeq}{\end{equation}}
\newcommand{\gcc}{{\rm~g\,cm}^{-3}}
\begin{document}
\title{Ionization Equilibrium and 
Equation of State of Hydrogen Plasmas
in Strong Magnetic Fields}
\author{A.~Y. Potekhin (a), G. Chabrier (b), Yu.~A. Shibanov (a),
J.~Ventura (c,d)}
\address
{(a) Ioffe Physical-Technical Institute,
     194021 St.-Petersburg, Russia\\
(b) CRAL,
     Ecole Normale Sup\'erieure de Lyon,
     69364 Lyon Cedex 07, France\\
(c) Department of Physics, University of Crete, 
               710\,03 Heraklion, Crete, Greece\\
(d) Institute of Electronic Structure and Laser, 
               FORTH, 711\,10 Heraklion, Crete, Greece
}
\maketitle
\begin{abstract}
We study
  hydrogen plasmas at magnetic fields $B\sim 10^{12}-10^{13}$~G,
densities $\rho\sim 10^{-3}-10^3{\rm~g~cm}^{-3}$
and temperatures $T\sim 10^{5.5}-10^{6.5}$~K,
  typical of photospheres of middle-aged cooling neutron stars.
We construct an analytical free energy model of the 
partially ionized plasma,
  including into consideration the decentred atomic states,
  which arise due to the thermal motion across the strong field.
  We show that these states,
   neglected in previous studies,
  may contribute appreciably into 
   thermodynamics of the outer atmospheric layers
  at $\rho \lesssim 1{\rm~g~cm}^{-3}$ and typical $B$ and $T$. 
  We take into account
  Coulomb non-ideality of the ionized component of the plasma
  affected by intense magnetic field.
  Ionization degree, occupancies and equation of state
  are calculated, and their dependences on the temperature, density and 
  magnetic field are studied.
\end{abstract}
\section{Introduction}
Magnetic fields $B\sim10^{12}-10^{13}$~G typical of 
isolated neutron stars qualitatively modify many 
physical properties of matter \cite{CanutoVentura,YaK}.
It was suggested that the outer layers of the neutron stars
may be composed of hydrogen at temperatures $T\sim10^{5.5}-10^{6.5}$~K
\cite{PCY}. Thus the study of hydrogen plasmas at such $B$ and $T$
is of great practical importance for astrophysics.
For studying the magnetized matter, 
Thomas-Fermi-like methods were used  
starting from 1970 \cite{Kadomtsev}
(see ref.\ \cite{Thor} for recent results and references).
It is well known, however, 
that they are not well suited for light elements.
Here we employ the free-energy minimization method.

The motion of charged particles in a magnetic field
is quantized into Landau orbitals.
The magnetic field is called {\it strongly quantizing\/}
if the free electrons populate mostly the ground Landau level \cite{YaK}.
This occurs when the electron cyclotron energy 
$\hbar\omega_c=\hbar eB/(m_e c)$ (where $\hbar$, $e$, $m_e$ and $c$
are the Planck constant, electron charge, electron mass and speed of light,
respectively)
exceeds both the thermal energy $k_B T$
and the electron Fermi energy $\epsilon_F$ --- that is for temperatures 
$T\ll T_B$ and densities $\rho<\rho_B$, where
\beq
   T_B=3.16\times 10^5\,\gamma{\rm~K} , 
\quad 
   \rho_B=0.809\,\gamma^{3/2}\gcc,
\quad
   \gamma\equiv{\hbar^3 B \over m_e^2 c e^3}={B\over2.35\times10^9\mbox{~G}}.
\label{strong-field}
\eeq
The atom in a {\it strong\/} magnetic field $\gamma\gg1$ 
is compressed in 
the transverse directions to the size of the ``magnetic length'':
$
   a_m=(\hbar c/eB)^{1/2}=a_0\,\gamma^{-1/2},
$
where $a_0=\hbar^2/(m_e e^2)$ is the Bohr radius.
The ground-state binding energy grows 
logarithmically with $B$
and exceeds the ground-state energy of the field-free atom 
by order of magnitude at $B\sim10^{12}$~G \cite{CanutoVentura}.
Ionization equilibrium of atoms
in strong magnetic fields has been first discussed 
in ref.\ \cite{Gnedin-etal}.
However, that pioneering work neglected modifications 
of the atomic properties caused by the thermal motion
of the atoms across the field.
These motional modifications arise from the coupling
between the centre-of-mass motion  
across the field and the relative electron-proton
motion.
These effects were appreciated
by Ventura et al. \cite{Ventura-etal}, 
but quantum-mechanical calculations of 
binding energies and wave functions
of hydrogen atoms in {\em any\/} states of motion
in the strong magnetic
fields have been carried out only recently \cite{P94}.

Lai and Salpeter \cite{LS97}
(see references therein for earlier work)
considered the ionization equilibrium
of strongly magnetized hydrogen using a crude approximation for 
binding energies of moving atoms which missed
the so-called {\it decentred states}
with a large electron-proton separation
\cite{P94}.
The same approximation was used in ref.\ \cite{Steinberg},
devoted to the low-density equation of state.
Here we employ new
fitting formulae to atomic energies and sizes \cite{P98}
based on the previous study \cite{P94}, valid for 
any state of atomic motion.
We construct an analytic model of the plasma free energy  
and derive and solve a generalized Saha equation. 

\section{Free Energy Model and Generalized Saha Equation}
\label{sect-model}
We consider a plasma consisting of $N_e$ electrons, $N_p=N_e$ protons, 
and $N_H$ hydrogen atoms
in a volume $V$,
and write the Helmholtz free energy as
$
    F = F_{\rm id} + F_{\rm ex},
$
where 
$
   F_{\rm id}=F_{\rm id}^{(e)}+F_{\rm id}^{(p)}
           +F_{\rm id}^{\rm neu}
$ 
is the sum of the ideal-gas free energies of the electrons,
protons, and neutral species, respectively,
and $F_{\rm ex}$ is the {\it excess\/} free energy.

For the ideal gas of electrons,
the pressure and number density are
\beq
   P_e = {k_B T\over\pi^{3/2} a_m^2\lambda_e}
     \sum_{N=0}^\infty g_{N} I_{1/2}(\beta\mu_N),
\quad
   n_e = {1\over2\pi^{3/2} a_m^2 \lambda_e}
     \sum_{N=0}^\infty g_{N} I_{-1/2}(\beta\mu_N),
\label{n_e}
\eeq
where $I_p(x)=\int_0^\infty t^p {\rm d}t /(e^{t-x}+1)$ 
is the Fermi integral, 
$\mu_N\equiv\mu_e-N\hbar\omega_c$, $\mu_e$ is the chemical potential,
$\beta\equiv(k_B T)^{-1}$, 
$\lambda_e\equiv\hbar\sqrt{2\pi\beta/m_e}$,
$g_{N\geq1}=2$, and $g_{N=0}=1$.
The free energy is given by 
$F_{\rm id}^{(e)}  = 
     \mu_e N_e - P_e V,$
where $\mu_e$ is found using an algorithm described in ref.\ \cite{PY}.

In the strongly quantizing regime,
the Fermi energy is $\epsilon_F = 2\pi^4 \hbar^2\,(a_m^2 n_e)^2 / m_e$,
which differs from the non-magnetic case 
by a factor $(4/3)^{2/3}(\rho/\rho_B)^{4/3}$.
Thus the degeneracy is strongly reduced at $\rho\ll\rho_B$.
Furthermore, in the non-degenerate regime ($k_B T \gg \epsilon_F$), 
we have
$
   F_{\rm id}^{(e)} = 
      N_e k_B T \left[ \ln(2\pi a_m^2\lambda_e n_e) - 1 \right].
$

For the protons, which are non-degenerate, we have
\beq
   \beta F_{\rm id}^{(p)}/N_p =
      \ln(2\pi a_m^2\lambda_p n_p)
    + \ln\left[1-\exp(-\beta\hbar\omega_{cp})\right]-1,
\label{Fp}
\eeq
where $\omega_{cp}=(m_e/m_p)\omega_c$ is the proton cyclotron frequency.
Here, for sake of brevity, we drop the zero-point
energy $\frac12\hbar\omega_{cp}$ and the spin energy
$\pm \frac14 g_p\hbar\omega_{cp}$, where $g_p=5.585$
is the proton spin gyromagnetic factor.
These terms are the same for free and bound protons.
Taking them into account yields an additive contribution:
$
  \Delta F = N_0 \{ \hbar\omega_{cp}/2
    - k_B T \ln[2\cosh(\beta g_p\hbar\omega_{cp}/4)] \},
$
where $N_0$ is the total number of protons (free and bound).
$\Delta F$ does not affect ionization equilibrium and pressure.

For the excess free energy of the ionized component,
a general fitting formula in the non-magnetic case is given 
in ref.\ \cite{CP98}.
It is known that 
thermodynamics of {\it classical} Coulomb plasmas is not affected
by the magnetic field, which, however, affects
the {\it quantum-mechanical} contributions to $F_{\rm ex}$.
These effects have been studied 
only in the low-temperature or low-density regimes
(e.g., ref.\ \cite{Steinberg} and references therein).
Here we use a scaling ($r_s^{\rm eff}=s r_s$) of the density parameter
$r_s=(4\pi n_e a_0^3/3)^{-1/3}$
at a fixed Coulomb parameter $\Gamma=\beta e^2/(a_0 r_s)$
 in the formulae of ref.\ \cite{CP98}.
The scaling is devised so as to reproduce
the low-density, high-temperature
results presented in ref.\ \cite{Steinberg}, as well 
as other known limiting cases.
For the contribution of electron-electron and electron-ion 
interactions in $F_{\rm ex}$, 
the scaling factors are
$
    s_{ee}=  (1+\theta_m/\theta_0)  /
           \left[  1+(\theta_m/\theta_0)\,\exp(-\theta_m^{-1}) f_1 \right]
$
and $s_{ie}=1/f_2^2,$
where  $\theta_0=2\,(9\pi/4)^{-2/3} r_s/\Gamma$ and 
$\theta_m=8\,\gamma^2 r_s^5/(9\pi^2\Gamma)$ are the non-magnetic and magnetic 
degeneracy parameters, respectively,
and the factors $f_1$ and $f_2$ (depending on $\beta\hbar\omega_c$)
are given in ref.\ \cite{Steinberg}.

The ideal-gas contribution of the magnetized atoms reads 
\beq
\beta F_{\rm id}^{(H)}=\sum_{s\nu}\int{\rm d}^2 K_\perp N_{s\nu}
(K_\perp)\left\{\ln\left[n_H\lambda_H^3w_{s\nu}(K_\perp)/Z_w\right]
-1\right\},
\eeq 
where
$s$ 
and   $\nu$ 
relate to electronic excitations,  
$N_{s\nu} = (\lambda_H/2\pi\hbar)^2
N_H w_{s\nu}{\rm e}^{\beta \chi_{s\nu}}/Z_w$ 
are the atomic occupancies 
per unit phase space of the transverse component $K_\perp$ 
of the pseudomomentum ${\bf K}$ which characterizes the atomic motion
in the magnetic field, 
 $w_{s\nu}(K_\perp)$ and $\chi_{s\nu}(K_\perp)$ 
are the occupation probabilities  
and binding energies 
of the moving atom, and
$Z_w=(\lambda_H/2\pi\hbar)^2
\sum_{s\nu}\int {\rm d^2}{K_\perp}
w_{s\nu}({K_\perp})\exp[\beta\chi_{s\nu}(K_\perp)]$ 
is the internal partition 
function. 

The contribution of atoms in the nonideal part $F_{\rm ex}$ of the
free energy is calculated in the
hard-sphere approximation using
the van der Waals one-fluid model by analogy with ref.\ \cite{P96}.
Its straightforward generalization to the magnetic case
involves the composite quantum number $\kappa=(s\nu K_\perp)$,
so that $\Sigma_\kappa$ includes now integration over $K_\perp$.
The hard-sphere diameters are set equal to
the effective atomic sizes $l_\kappa$
given in \cite{P98}.
The occupation probabilities are then given by
formulae derived in ref.\ \cite{P96}, 
extended to the magnetic case.

Our model is valid as long as the formation of molecules may be neglected.
In order to quantify the range of validity,
we estimate the abundance of H$_2$ molecules following ref.\ \cite{LS97},
but with inclusion of the non-ideal effects.

Minimization of the free energy yields the ionization equilibrium
(generalized Saha) equation:
\beq
    n_H =  n_p n_e                                           
   (\lambda_p\lambda_e /\lambda_H^3) (2\pi a_m^2)^2
    \, \left[1-\exp(-\beta\hbar\omega_{cp})\right] \, 
    Z_w \exp(\Lambda),
\eeq
where 
$
   \Lambda=
        \beta \mu_e  - \ln(2\pi a_m^2\lambda_e n_e)
    + \beta\, \partial \mu_e/\partial\ln n_e - 
        \partial P_e / \partial n_e
$
takes into account effects of 
electron degeneracy and population of excited Landau levels.

\section{Results and Discussion}
\begin{figure}[h]
\begin{center}
\epsfig{file=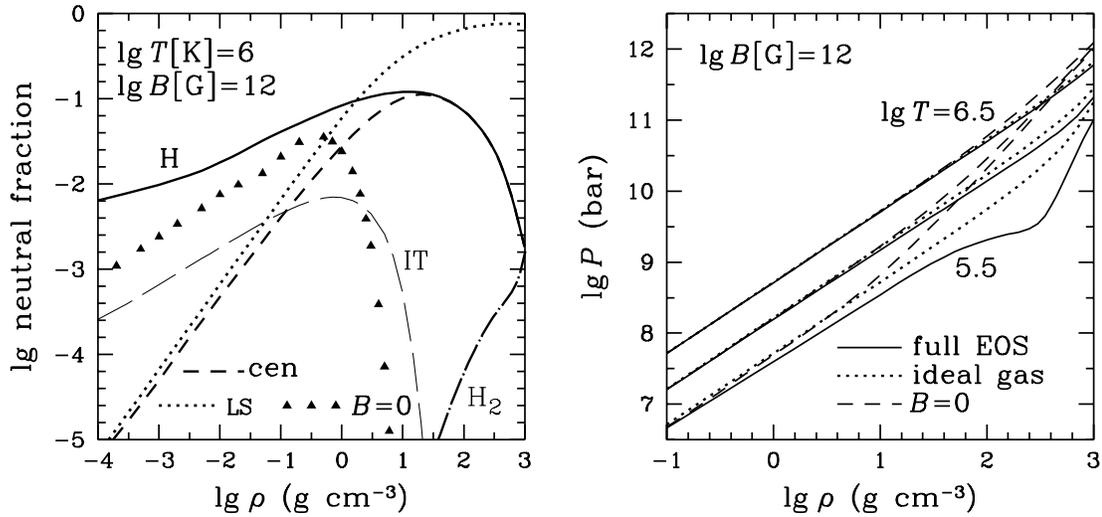,width=15.5cm}
\end{center}
\caption{Left panel: Non-ionized fraction of atoms in any states 
(solid line),
atoms in the centred states (short-dashed line),
molecules (dash-dot line), and the weakly perturbed atoms contributing
to the optics (long-dashed line),
compared with the non-magnetic case (triangles) 
and the approximation \protect\cite{LS97}
(dotted line).
Right panel: Pressure isotherms of magnetized hydrogen plasma
(solid lines) at $\lg T \mbox{[K]} = 5.5$, 6.0 and 6.5,
compared with the non-magnetic case (dashed lines)
and with the ideal magnetized plasma (dotted lines).}
\label{fig1}
\end{figure}

Figure \ref{fig1} shows selected results obtained for $B=10^{12}$~G.
The left panel shows the neutral fraction of atoms $f_H=N_H/N_0$ 
and molecules $f_{H2}=2N_{H2}/N_0$ at $T=10^6$~K.
For comparison, we plot the fraction of atoms in the centred states, 
$f_H$ according to ref.\ \cite{LS97} 
and $f_H$ in the non-magnetic case.
Long dashes display the fraction of atoms that satisfy 
the Inglis--Teller (IT) criterion and thus can be identified
in optical spectra of the plasma. The IT fraction is estimated
according to the formula 
$n_\kappa^{\rm IT} \sim n_\kappa
\exp[-n_p (4l_\kappa)^3]$
(cf. Eq.~(31) of ref.\ \cite{P96}). 
We can see that (a) the strong magnetic field increases the non-ionized
fraction and shifts the region of pressure ionization
to much higher $\rho$ 
(compare the solid line and triangles in the left panel of Fig.~\ref{fig1}), 
(b) the approximation of ref.\ \cite{LS97} reproduces
only the abundance of the centred atoms at low density
and fails at high density where the pressure-ionization 
effects are important,
and (c) at the low density, the decentred atomic states 
are significantly populated. 

The right panel demonstrates the equation of state,
which is seen to be much softer than (a) in the 
non-magnetic case (mainly because of the electron degeneracy
``taken away'' by the strongly quantizing field,
but also due to the increased neutral fraction)
and (b) in the magnetic but ideal 
proton-electron plasma (because the Coulomb interactions
yield negative contribution to the pressure).

The obtained results are used for modelling neutron-star atmospheres.
In particular, the IT fraction of atoms,
multiplied by the absorption cross sections 
calculated in ref.\ \cite{PP97},
determines an atomic contribution to atmospheric opacities.
Preliminary calculations of the opacities, 
carried out with a simplified $F_{\rm ex}$, were presented
in ref.\ \cite{PSV}. 
The more elaborated model of the plasma described here
confirms qualitative results of that work.
An important conclusion is that the bound
species contribute significantly to the absorption
at $B=10^{12}-10^{13}$~G, even at relatively high $T\sim10^6$~K.

\begin{acknowledgements}
The work was supported in part by the grants
RFBR 96-02-16870a, DFG--RFBR 96-02-00177G, and INTAS 96-0542.
A.Y.P. acknowlegdes 
a visiting professorship in the theoretical astrophysics group
of the Ecole Normale Sup\'erieure de Lyon.
The participation of A.Y.P. in the PNP-9 Workshop 
has been supported in part by Deutsche Forschungsgemeinschaft 
and Russian Foundation for Basic Research.
\end{acknowledgements}
\vspace*{-1cm}
\begin{received}Received 1 October 1998
\end{received}
\end{document}